\documentclass[a4paper,10pt]{article}
\usepackage{amsmath,amsthm,latexsym,amssymb,amsfonts}

\newcommand{\h}{\hat{\rm h}}

\newcommand{\tg}{{\rm tg}}
\newcommand{\A}{{\tilde A}}
\newcommand{\B}{{\tilde B}}
\newcommand{\bra}{\langle}
\newcommand{\ket}{\rangle}
\newcommand{\Z}{{\mathbb Z}}
\newcommand{\R}{{\mathbb R}}
\newcommand{\C}{{\mathbb C}}

\newcommand{\hilb}{{\mathcal H}}
\newcommand{\F}{{\mathcal F}}
\newcommand{\spane}{{\rm span}}

\newcommand{\ch}{{\rm ch}}
\newtheorem{lm}{Lemma}
\newtheorem{thm}[lm]{Theorem}

\begin{document}

\title{The flat FRW model in LQC: the self-adjointness}
\author{ Wojciech Kami\'nski\thanks{wkaminsk@fuw.edu.pl}\ \
and\ \
        Jerzy Lewandowski\thanks{lewand@fuw.edu.pl}}
\date{\it 1. Instytut Fizyki Teoretycznej,
Uniwersytet Warszawski, ul. Ho\.{z}a 69, 00-681 Warszawa, Poland\\
2. Physics Department, 104 Davey, Penn State, University Park, PA 1602, USA \\
[.5cm]} \maketitle
\begin{abstract}
The flat FRW model coupled to the massless scalar field according to
the improved, background scale independent version of Ashtekar,
Pawlowski and Singh \cite{APS} is considered. The core of the theory
is addressed directly:  the APS construction of the quantum
Hamiltonian is analyzed under the assumption that cosmological
constant $\Lambda\le 0$. We prove the essential self-adjointness of
the operator whose square-root defines in \cite{APS} the quantum
Hamiltonian operator and therefore, provide the explicit definition.
If $\Lambda<0$, then  the spectrum is discrete. In the $\Lambda=0$
case, the essential and absolutely continues spectra of the operator
are derived. The latter operator is related in the unitary way to
the absolutely continuous part of the Quantum Mechanics operator
$a(-\frac{\partial^2}{\partial y^2} - \frac{b}{{\rm
cosh}^2\hat{y}})$  ($a,b>0$ being some constants) plus a trace class
operator.
\end{abstract}

\section{Introduction} Loop Quantum Cosmology \cite{APS,LQCearlier} is a
developing topic, grows into a well established theory and provides
both qualitative and quantitative results which change our believes
about the origins of the Universe \cite{APS}.  The theory has
achieved the point when it deserves study of  the exact properties
of the quantum operators describing the evolution of the quantum
space-time geometry. The work in that direction has been already
initiated in \cite{k=1} (see SKL), on the occasion of deriving the
gravitational part of the quantum scalar constraint operator in the
case of the closed Freedman Robertson Walker (FRW) model.

In this paper we consider the flat FRW model coupled to the massless
scalar field according to the improved, background scale independent
version of Ashtekar, Paw{\l}owski and Singh \cite{APS}. We are
assuming that the cosmological constant
$$\Lambda\le 0.$$
 We go directly to core of
the theory and analyze the APS construction of the quantum
Hamiltonian. We study the operator whose square-root defines in
\cite{APS} the quantum Hamiltonian operator and show it is
essentially self-adjoint. That result provides explicit definition
of the quantum Hamiltonian. The same definition was already
correctly anticipated in the numerical calculations in \cite{APS} in
the case $\Lambda=0$. Here, we provide an exact proof.  We derive
also the essential and absolutely continues spectra of the operator.
Moreover, we construct a transform that maps the operator in a
unitary fashion into  the very well known in Quantum Mechanics
operator $a(-\frac{\partial^2}{\partial y^2} - \frac{b}{{\rm
cosh}^2\hat{y}})$ ($a,b>0$ being some constants) plus a trace class
operator. In the case $\Lambda<0$ we show that that the operator has
purely discrete spectrum.
\medskip

Below, we introduce all those elements of the APS model which will
be needed to understand our results and derivation. For the
physics of the model and all the technical details which are not
relevant  in what follows, the reader is referred to the original
paper \cite{APS}.

The kinematical Hilbert space for the gravitational degrees of
freedom in the FRW model is
\begin{equation}
\hilb_{\rm gr}\ =\ \overline{\spane\{\ |v\ket\ :\ v\in\R  \ \} }
\end{equation}
with the scalar product
\begin{equation}
\bra v|v'\ket\ =\ \begin{cases}1,&  {\rm if}\ \ v=v'\\0,&\ \  {\rm
otherwise},\end{cases}
\end{equation}
The gravitational degrees of freedom are represented by the quantum
volume operator $\hat{v}$,
\begin{equation}
\hat{v}|v\ket\ =\  v |v\ket,
\end{equation}
defined in the domain
\begin{equation}\label{D}
D\ =\ \spane\{\ |v\ket\ :\ v\in\R  \ \}
\end{equation}
 and by the improved holonomy
operator
\begin{equation}\label{hnu}
{\h}_\nu |v\ket\ =\ |v+\nu\ket,
\end{equation}
defined in the entire Hilbert space $\hilb_{\rm gr}$.

Given a function $Q:\R\rightarrow \R$, the corresponding
multiplication operator will be denoted by $Q(\hat{v})$,
\begin{equation}\label{multop}
Q(\hat{v})|v\ket\ =\ Q(v)|v\ket.
\end{equation}

The quantum volume  operator measures (modulo a constant factor
\cite{APS}) the physical volume of some cubic cell fixed in a
homogenous 3-slice of space-time, whereas the improved holonomy
operator contains the information about the extrinsic curvature of
the 3-slice.
\bigskip

The kinematical Hilbert space of the homogeneous scalar field is
just
\begin{equation}
\hilb_{\rm sc}\ =\ L^2(\R),
\end{equation}
the measure being the Lebesgue one. The field operator
$\hat{\phi}$ and the canonically conjugate momentum $\hat{p}_\phi$
are defined as in the Shroedinger quantum mechanics,
\begin{equation}
(\hat{\phi}\psi)(\phi)\ =\ \phi\psi(\phi),\ \ \
\left(\hat{p}_\phi\psi\right)(\phi)\ =\
- i\frac{\partial}{\partial\phi}\psi(\phi).
\end{equation}
\bigskip

The total kinematical Hilbert space is
\begin{equation}
\hilb\ =\ \hilb_{\rm gr}\otimes\hilb_{\rm sc}.
\end{equation}
In this model all the Einstein-Klein-Gordon constraints are solved
classically except the scalar constraint which takes the form of
the following quantum scalar constraint operator
\begin{equation}
\hat{C}\ =\ {B(\hat{v})}\otimes\hat{p}_\phi^2\  -\ \hat{C}_{\rm gr}\otimes {\rm id},
\end{equation}
defined in the domain
$$D(\hat{C})\ =\ \spane\{\ |v\ket\ :\ v\in\R \} \otimes {\rm C}^\infty_0(\R)$$
where  $B(\hat{v})$ is proportional to the quantum operator coming
from the quantization of the classical inverse volume, namely
\begin{equation}\label{B}
B(v)\ =\ \frac{27}{8}|v|\left||v+1|^{1/3}-|v-1|^{1/3}\right|^3,
\end{equation}
and  $\hat{C}_{\rm gr}$ is the gravitational field part of the
scalar constraint, namely
\begin{align}
\hat{C}_{\rm gr}\ &=\ -\h_2 A(\hat{v})\h_2\ -\
\h_{-2}A(\hat{v})\h_{-2}\ +\
A(\hat{v}+2)\ +\ A(\hat{v}-2)\ +\ \tilde{\Lambda}|\hat{v}|,\label{Cgr}\\
A(v)\ &=\ \frac{3 \pi  G}{8}|v| \cdot \left|(|v+1|-|v-1|)
\right|,\label{A}
\end{align}
where $\tilde{\Lambda}$ is proportional to the cosmological constant
$\Lambda$ and the cube of the Planck length $\ell$,
 \begin{equation}
\tilde{\Lambda}\ =\
-\left(\frac{4\pi\gamma}{3}\right)^{\frac{3}{2}}\frac{3\sqrt{3\sqrt{3}}}{2\sqrt{2}}
\ell^3\Lambda.
\end{equation}
In this paper we are considering the case
\begin{equation}
\Lambda\ \le \ 0.
\end{equation}

\bigskip

The scalar constraint operator $\hat{C}$ preserves every subspace
$D_\epsilon \otimes {\rm C}^\infty_0(\R)$ defined by
\begin{equation}\label{Depsilon}
D_{\epsilon}\ =\ {\spane\{\  |v\ket\ :\ v=4n+\epsilon, n\in\Z \}},
\end{equation}
where $\epsilon\in [0,4)$.  There is an orthogonal decomposition
\begin{equation}
\hilb_{\rm gr}\ =\ \overline{\oplus_{\epsilon\in
[0,4)}D_{\epsilon}}.
\end{equation}
\bigskip
The subspaces $\hilb_{\rm \epsilon}\otimes \hilb_{sc}$ where
\begin{equation}\label{hilbepsilon}\hilb_{\epsilon} \ =\
\overline{D_\epsilon},
\end{equation}
 are sometimes considered
analogs of "super-selection sectors", meaning that the
representation of the given quantum theory is reducible and should
be reduced to one of the subspaces.

In order to ``solve'' the quantum scalar constraint, and use the
solutions to define the physical Hilbert space, one considers the
restriction of the operator $\hat{C}$ to any of the subspaces
$\hilb_{\epsilon}\otimes \hilb_{\rm sc}$ and tries to give a
meaning to the quantum counterpart of the scalar constraint
equation, namely to
\begin{equation}\label{C1}
``\hat{C}|\psi\ket\ =\ 0''.
\end{equation}
The problem is that there are not sufficiently many solutions of
that equation in  $\hilb_{\rm sc}\otimes \hilb_{\epsilon}$.
Moreover,  our experience in the standard quantization shows, that
unless  the gauge group generated by the constraints is compact, the
physical solutions to the quantum constraint are
``non-normalizable'', that is they belong to a new Hilbert space,
constructed from the starting kinematical Hilbert space.

In this paper we focus on the quantization scheme introduced and
successfully applied by \cite{APS}.

The starting point is replacing the heuristic formula (\ref{C1}) by
another, also heuristic formula
\begin{equation}\label{C2}
"\big({\rm id}\otimes \hat{p}_\phi\  -\ \sqrt{B(\hat{v})^{-1}
\hat{C}_{\rm gr}}\otimes {\rm id}\big)|\psi\ket\ =\ 0".
\end{equation}
The difference between (\ref{C1}) and (\ref{C2}) consists in the two steps:
\begin{itemize}
\item[$i)$] going from the operator $\hat{C}$ to $B(\hat{v})^{-1}\hat{C}$;
\item[$ii)$]
``selecting the positive frequency modes'' of the $\hat{p}_{\phi}$ operator,
meaning using the spectral decomposition of the Hilbert space defined by
$\hat{p}_{\phi}$ and restricting to the non-negative part of the spectrum.
\end{itemize}

To give (\ref{C2}) the exact meaning, APS  provide the vector space
$D_\epsilon$  with a new scalar product $\langle\cdot,\cdot\rangle'$
such that
\begin{equation}\label{scprod'}
\bra v_1|v_2\ket'\ :=\ \begin{cases}B(v_1),&  {\rm if}\ \
v_1=v_2\\0,&\ \ {\rm otherwise}\end{cases}
\end{equation}
The resulting Hilbert space $\hilb_\epsilon'$, that is the
completion of $D_\epsilon$ in the new scalar product, is promoted
to be the physical Hilbert space. We will be also assuming the
generic case
$$\epsilon\ \not=\ 0$$
when the restriction of the $B(\hat{v})$ operator is invertible.

 The operator $B(\hat{v})^{-1}\hat{C}_{\rm gr}$
preserves $D_\epsilon$, can be written in a manifestly positive
definite form and is symmetric (due to the modification in the
scalar product). It follows that it admits a self-adjoint
extension, say $\hat{\theta}$. It is used in \cite{APS} to give
the exact meaning to (\ref{C2}). Solutions of (\ref{C2}) are
defined to be all the maps
$$ \R\ni \phi\mapsto U_{\phi}\Psi\in \hilb_\epsilon'$$
such that
$$U_{\phi} \ =\ \exp\left(i\phi \sqrt{\hat{\theta}}\right), \ \ {\rm and}\ \ \Psi\in\hilb_\epsilon'.$$
However, the unitary map $U_{\phi}$ depends on the choice of the
self-adjoint extension $\hat{\theta}$, unless the extension is
unique. APS circumvent that problem, by declaring a choice of the
so called Friedrichs extension, a mathematically distinguished
extension which exists and is unique. That is a natural choice and
makes the APS model complete. The open question was, what form
that extension took in the case at hand, and whether or not there
were possible other self-adjoint extensions that could {\it a
priori} have different physical properties. Those issues gave the
motivation for our current paper.

\section{The results} In this section we present our results. The case
$\Lambda<0$ is simple enough to outline the proof. In the case
$\Lambda=0$ the sketch of the proof presented in this section will
be followed by the detailed derivation in the subsequent sections.

The starting point for our work is the operator
$$B(\hat{v})^{-1}\hat{C}_{\rm gr}$$
introduced above (see (\ref{Cgr}), (\ref{B})), defined in the domain
$$D(B(\hat{v})^{-1}\hat{C}_{\rm gr})\ =\ D_\epsilon$$
in the Hilbert space $\hilb_\epsilon'$ (see
(\ref{Depsilon}),(\ref{scprod'})). Our first step is using a unitary
map
\begin{align}
B(\hat{v})^{\frac{1}{2}}\ :\ \hilb_\epsilon'\ &\rightarrow\ \hilb_\epsilon,\\
|v\ket\ &\mapsto\ \sqrt{B(v)}|v\ket\nonumber
\end{align}
which  maps the physical Hilbert space back into the Hilbert space
(\ref{hilbepsilon}). The operator in question is carried into the
following operator
\begin{equation}\label{C}
 {B(\hat{v})}^{-\frac{1}{2}}\circ
\hat{C}_{\rm gr}\circ {B(\hat{v})}^{-\frac{1}{2}}
\end{equation}
defined in the domain $D_{\epsilon}$.

In order to show to what extent our results do not depend on the
definition of the functions $A$ and $B$, we consider in this work
the following generalization of the operator $\hat{C}_{\rm gr}$:
\begin{align}\label{HAPS}
{H}_{\rm APS}\ &=  {\B(\hat{v})}^{-\frac{1}{2}}\big(-\h_2
\A(\hat{v})\h_2\ -\ \h_{-2} \A(\hat{v})\h_{-2}\ \nonumber\\ &+\
\A(\hat{v}+2)\ +\ \A(\hat{v}-2)\ +\ \tilde{\Lambda}|\hat{v}|\big)
{\B(\hat{v})}^{-\frac{1}{2}},
\end{align}
defined in the domain
$$D(H_{\rm APS})\ =\ D_{\epsilon}$$
(\ref{Depsilon}) in the Hilbert space $\hilb_{\epsilon}$
(\ref{hilbepsilon}),
where the operator $\h_2$ is defined in (\ref{hnu}),  and about
the functions $\A,\B:\R\rightarrow \R$ we are assuming only what
follows:
\begin{align} \A(v)\ &=\ \frac{3}{4}\pi G |v| + o_0(v),\label{A'}\\
   \B(v) \ &=\ \frac{1}{|v|}(1+\frac{\alpha}{|v|^2} +
O(\frac{1}{v^4})),\label{B'}
\end{align}
where $o_0$ has a compact support. The functions  (\ref{A}) and
(\ref{B}) are a special case of   (\ref{A'}) and (\ref{B'}) with
\begin{equation}
\alpha\ = \ \frac{10}{18}.
\end{equation}

The operator  $H_{\rm APS}$ is manifestly symmetric. It is also
positive definite provided $A(v)\ge 0$ for every $v=4n+\epsilon$,
$n\in\Z$ and given $\epsilon$. To see the latter property it is
enough to write the operator in the following form:
$$H_{\rm APS}\ =\ \tilde{B}(\hat{v})^{-\frac{1}{2}} \frac{(\h_2-\h_{-2})}{i}
\tilde{A}(\hat{v})\frac{(\h_2-\h_{-2})}{i}
\tilde{B}(\hat{v})^{-\frac{1}{2}}\ +\
\tilde{\Lambda}|\hat{v}|\tilde{B}(\hat{v})^{-1}.$$

We study separately the cases  $\tilde{\Lambda}> 0$ (that is the
cosmological constant $\Lambda<0$), and, respectively,
$\tilde{\Lambda}=0$ (that is $\Lambda=0$).

\begin{thm}  Suppose
\begin{equation}
\tilde{\Lambda}\ >\ 0.
\end{equation}
Then, for arbitrary $\epsilon\in(0,4\pi)$, the operator $H_{\rm
APS}$ (\ref{HAPS}) has the following properties:
\begin{itemize}
\item[$i$)] it is  is essentially self
adjoint;
\item[$ii$)] it has discrete spectrum;
\item[$iii$)]  the spectrum  is
contained in $(0,\infty)$
\item[$iv$)] for arbitrary $E>0$ the dimension of the Hilbert subspace
spanned by the eigenvectors such that the corresponding eigenvalue
$\lambda$ satisfies
$$\lambda < E$$
is less or equal the number of elements of the set
$$\{n\in\Z \ :\ \tilde{\Lambda}\frac{|4n+\epsilon|}{\tilde{B}(4n+\epsilon)}\
< E\} $$
\end{itemize}
\end{thm}
\bigskip

Technically, the theorem and proof are similar to those presented in
\cite{k=1} (see SKL) concerning the operator $\hat{C}_{\rm gr}$ in
the $k=1$ case. Therefore we just outline the proof. To begin with,
we write the operator $H_{\rm APS}$ in the following form
\begin{align}
H_{\rm APS}\ =\
-\h_2&\frac{\A(\hat{v})}{\sqrt{\B(\hat{v}+2)\B(\hat{v}-2)}}\h_2
-\h_{-2}\frac{\A(\hat{v})}{\sqrt{\B(\hat{v}+2)\B(\hat{v}-2)}}\h_{-2} +
\nonumber\\
&+\frac{\A(\hat{v}+2)+\A(\hat{v}-2)+
\tilde{\Lambda}|\hat{v}|}{\B(\hat{v})}\ .
\end{align}
Then we split the operator into two operators $H_0$ and $H_1$
defined in the same domain, namely
\begin{align}
H_{\rm APS}\ &=\ H_1\ +\ H_0\\
H_1\ &=\
-\h_2\frac{\A(\hat{v})}{\sqrt{\B(\hat{v}+2)\B(\hat{v}-2)}}\h_2
-\h_{-2}\frac{\A(\hat{v})}{\sqrt{\B(\hat{v}+2)\B(\hat{v}-2)}}\h_{-2}\\
H_0\ &=\ \frac{\A(\hat{v}+2)+\A(\hat{v}-2)+\
\tilde{\Lambda}|\hat{v}|}{\B(\hat{v})}\ .
\end{align}
Note that the operator $H_0$ is essentially self-adjoint. It is not
hard to derive the following inequality for some constant $\beta'>0$
and arbitrary $\psi\in D_\epsilon$,
\begin{equation}
\|H_1\psi\|^2\ \le\ \|H_0\psi\|^2\ +\ \beta'\|\psi\|^2,
\end{equation}
For that purpose we use the expansion and expand the functions:
\begin{align}
\frac{\A({v})}{\sqrt{\B({v}+2)\B({v}-2)}} \ &=\
\frac{3\pi G}{4}\big({v}^2 - 2-\alpha \big) + O(\frac{1}{v^2})\label{exp1}\\
\frac{|v|}{\B({v})}\ &=\   \big( {v}^2 - \alpha \big) +
O(\frac{1}{v^2})\label{exp2},
\end{align}
and the fact, that the operator  $f(\hat{v})$ for every function
$f:\{v\ :\ v=4n+\epsilon,\ n\in\Z\}\rightarrow\R$ of a compact
support is bounded. The inequality shows the self-adjointness. The
properties of the spectrum follow from the inequality
$$ H_{\rm APS}\ \ge\
\tilde{\Lambda}|\hat{v}|\tilde{B}(\hat{v})^{-1}.$$
(see \cite{k=1}, SKL).
\bigskip

\begin{thm}
Suppose
\begin{equation}
\tilde{\Lambda}\ =\ 0.
\end{equation}
Then, for arbitrary $\epsilon\in(0,4\pi)$, the operator $H_{\rm
APS}$ (\ref{HAPS}) has the following properties:
\begin{itemize}
\item[$i$)] it is essentially self-adjoint;
\item[$ii$)] the absolutely continuous spectrum is $[0,\infty)$
\item[$iii)$] the essential spectrum is $[0,\infty)$.
\item[$iv$] the absolutely continues part of the operator $H_{\rm APS}$ is
unitarily equivalent to the operator $C^\infty_0(\R)\ni f\mapsto
-f''\in C^\infty_0(\R)$ in the Hilbert space $L^2(\R)$.
\end{itemize}
\end{thm}

\bigskip

For those readers who do not have much experience with the terms
used above\footnote{The exact definitions can be found in
\cite{Simon}. We quickly recall , that the essential spectrum of a
self-adjoint operator  $H$ is a complement of $\lambda\in \C$ such
that, $H-\lambda$ is invertible up to a compact operator. The
absolutely continues part of the operator is, vaguely speaking,
restriction to the subspace of vectors which spectral measure is
absolutely continuous with respect to the Lebesgue measure.
Absolutely continuous spectrum is the spectrum of this restricted
operator. This notion is strongly related to the scattering theory.
} we describe now
our result without referring to sophisticated elements of the
Hilbert space theory. What we do, and what is presented in detail in
the following sections, is we just construct a unitary
transformation which significantly simplifies the operator in
question modulo a trace class operator (that is an operator of a
finite trace). Then, we refer to the known properties of the simpler
operator.

More specifically, consider the operator given by the first two
terms of the expansion (\ref{exp1}, \ref{exp2}), namely
\begin{equation}
{H}_{\rm APS}'\ =\ \frac{3\pi G}{4}\big( -\h_2(\hat{v}^2 - 2-\alpha
)\h_2 - \h_{-2}(\hat{v}^2 -2-\alpha)\h_{-2} + 2\hat{v}^2 - 2\alpha
\big).
\end{equation}
It is not hard to check   that the difference between the operators
${H}_{\rm APS}$ and ${H}_{\rm APS}'$ is the following finite sum
\[
\h_2 g_1(\hat{v})\h_2+g_2(\hat{v})+\h_{-2} g_1(\hat{v})\h_{-2}
\]
where each of the functions $g_i$, $i=1,2$  vanishes in infinity
at least as fast as $v^{-2}$. Hence the sum is a compact, trace
class operator.

In the consequence \cite{Kato}, the operator $H_{\rm APS}$ is
essentially self adjoint if and only if $H_{\rm APS}'$ is. If the
operators are self adjoint, then their  spectra  mentioned in
Theorem  (the essential, and respectively,  absolutely continues
spectrum) are the same,  and also absolutely continues parts of
the operators are unitarily equivalent.

Therefore, we continue by studying the operator ${H}_{\rm APS}'$.
As it was mentioned in Sec. 1,  it turns out  there is a unitary
map which carries some very well known  operator in the standard
Quantum Mechanics into our ${H}_{\rm APS}'$. Before discussing the
properties of that map, we need the following short preparation.
In order to compare a given symmetric operator $X$ defined in a
domain $D(X)$, with another, self adjoint operator, one provides
$D(X)$ with the graph norm $\|\cdot\|_X$ defined as follows,
$$\|\cdot\|_X^2=\|\cdot\|^2+\|X\cdot\|^2.$$ Next, one considers
the completion $D(\bar{X})$ of $D(X)$ in the graph norm, and finally
the continues extension $\bar{X}$, of $X$ in $D(\bar{X})$ with
respect to the graph norm, so called closure of $X$. Therefore, we
have the closure $\overline{{H}_{\rm APS}'}$ of the operator
${H}_{\rm APS}'$ defined in the domain $D(\overline{H_{\rm APS}'}).
$\footnote{It is worth to be noted that if two operators are
unitarily equivalent the same is true for their closures, whereas
reverse statement is incorrect.}

Subject to the comparison, the second operator is ${H}_{\rm ch}:{\rm
C}^\infty_0(\R)\rightarrow {\rm C}^\infty_0(\R)$  defined in the
Hilbert space L$^2(\R)$, whose action on every  $f\in{\rm
C}^\infty_0(\R)$ is defined in the following way
\begin{equation}
\left({H}_{\rm ch}f\right)(y)\ = \left(-\frac{d^2}{dy^2} -
\frac{\alpha+1}{\ch^2(2{y})}\right)f(y).
\end{equation}

We  construct a unitary map $U:L^2(\R)\ \rightarrow \
\hilb_{\epsilon}$ which has the following two properties:
\begin{itemize}
\item[$i)$] $U$ carries ${\rm C}^\infty_0(\R)$ into
$D(\overline{H'_{\rm APS}})\subset\hilb_\epsilon$,
\item[$ii)$] The operator $\overline{H_{\rm APS}'}$ restricted
to the image $U({\rm C}^\infty_0(\R))$ satisfies
\begin{equation}\label{U}
U{H_{\rm ch}}U^{-1}\ = \frac{1}{3\pi G} \overline{H_{\rm APS}'}.
\end{equation}
\end{itemize}

Now,  the properties of operators such as $H_{\rm ch}$ are very well
known \cite{Simon}. In particular the operator is essentially self
adjoint. This is sufficient to conclude that the operator $H_{\rm
APS}'$ and in the consequence $H_{\rm APS}$ are essentially
self-adjoint. The essential spectrum of $H_{\rm ch}$ is $[0,\infty)$
as well as the absolutely continues spectrum. Those features are
preserved by every unitary map and trace class perturbation.
Finally, it is known that the absolutely continues part  of the
operator $H_{\rm ch}$ is unitarily equivalent to the part
$-\frac{d^2}{dy^2}$ of the operator. However, that does not exclude
existence of a point spectrum or singular continuous one. The whole
spectrum is also equal to $[0,\infty)$ due to the positive
definiteness.
\bigskip

What we are left with is the construction of the unitary map
(\ref{U}). It will be performed in the next two sections.

\section{The Fourier Transform} We define now a unitary map
\[
\F: \hilb_{\epsilon}\ \rightarrow\ L^2([0,\ 2\pi]),
\]
by the following formula
\begin{align}
\psi=\sum_{n\in\Z}\psi(4n+\epsilon)|4n+\epsilon\ket\ &\mapsto \F{\psi}\\
\F{\psi}(x)\ &=\ \sum_{n\in\Z} \psi(4n+\epsilon)e^{i(n+\frac{\epsilon}{4})x}
\end{align}

The operator $\h_{4}$ is transformed into the multiplication operator
$e^{i\hat{x}}$ (we use the convention analogous to (\ref{multop})),
\begin{equation}
(e^{i\hat{x}}{f})(x)\ =\ e^{ix}{f}(x).
\end{equation}

The operator $\frac{\hat{v}}{4}$ is transformed into the derivative
$-i\frac{\partial}{\partial x}$ of a domain, however, different then
the usual, namely:
\[
D(-i\frac{\partial}{\partial x})\ =\ \{f\in L^2([0,2\pi])\ : f
{\rm\  is\ absolutely\ continuous},\ f(2\pi)=e^{\frac{1}{2}i\pi\epsilon}f(0)\}.
\]
The operator generates a unitary group
\[
U_t (f)(x)=\left\{
\begin{array}{ll}
f(x+t) &, x+t\leq 2\pi\\
f(x+t-2\pi)e^{\frac{1}{2}i\pi\epsilon} &,x+t>2\pi
\end{array}
\right.
\]
In this section it is convenient to write the operator ${H}_{\rm
APS}'$ in the following form
\begin{align}
{H}_{\rm APS}'\ &=\ \frac{3\pi G}{4} [ -\left(\h_4
+\h_{-4}-2\right)\hat{v}^2 -4\left(\h_4-\h_{-4}\right)
\hat{v}\nonumber\\ &+\left(\alpha-2\right)\left(\h_4+\h_{-4}\right)
-2\alpha ].
\end{align}
The operator ${H}_{\rm APS}'$ is mapped by $\F$ into the following
operator
\begin{align}\label{Hbis}
{H''}_{\rm APS}\ &=\ 3\pi G [\ 4\left(\h_4 +\h_{-4}-2\right)
\frac{d^2}{dx^2}
+4i\left(\h_4-\h_{-4}\right)\frac{d}{dx}\nonumber\\&+
\frac{\alpha-2}{4} \left(\h_4 +\h_{-4}\right) -\frac{\alpha}{2}\ ]
\end{align}
%
whose natural domain $D({H''}_{\rm APS})$ is the image of the
domain of the closure of the operator ${H}_{\rm APS}'$. We will
characterize now a part of $D({H''}_{\rm APS})$ that will be
particularly useful in our paper. From the very definition each
function of the form
\[
g(x)=\sum_{k=n_0}^{n_1} a_k e^{i(k+\frac{\epsilon}{4})x}
\]
belongs to the domain $D({H''}_{\rm APS})$. The set of function
interesting from the point of view of this paper is
\begin{equation}\label{thedomain}
{\rm C}_{0,2\pi}^\infty([0,2\pi])\ :=\ \{f\in {\rm C}^\infty([0,2pi]\ :\
f\ {\rm vanishes\ in\ a\ neighborhood\ of\ 0\ and\ 2\pi} \}.
\end{equation}
The key result of this subsection is the following:
\begin{lm}
$${\rm C}_{0,2\pi}^\infty([0,2\pi]) \subset\
D(\overline{{H''}_{\rm APS}}) $$
\end{lm}
\begin{proof}
Writing the operator ${H''}_{\rm APS}$ in the following form
\begin{equation}
{H''}_{\rm APS}\ =\ 3\pi G \left(
-16\sin^2(\frac{\hat{x}}{2})\frac{d^2}{dx^2} -8\sin(\hat{x})
\frac{d}{dx}+ \frac{\alpha-2}{2} \cos(\hat{x})
-\frac{\alpha}{2}\right)\nonumber
\end{equation}
we can see an estimate:
\begin{align}\label{|Hf|}
|({H''}_{\rm APS}f)(x)|\ &\leq\ 3\pi G[\
16\left|\frac{d^2}{dx^2}f(x)\right|+
4\left|\frac{d}{dx}f(x)\right|\nonumber\\
&+ {\rm max}(\left|\frac{\alpha-2}{2} +\frac{\alpha}{2}\right|,
\left|\frac{\alpha-2}{2} -\frac{\alpha}{2}\right| )|f(x)|\ ].
\end{align}
 That gives us an estimate on the norms of the form
\begin{equation}
\lVert f\rVert^2+\lVert {H''}_{\rm APS} f\rVert^2\   \leq\ \beta
\lVert f''\rVert^2 +\gamma \lVert f'\rVert^2+ \chi \lVert
f\rVert^2, \label{norms}
\end{equation}
where $\beta, \gamma, \chi > 0$ are some constants one can calculate from
(\ref{|Hf|}) applying the inequality
$$\left(\left|a\right|+|\left|b\right|+\left|c\right|\right)^2\ \le
\ 3 \left(\left|a\right|^2+|\left|b\right|^2+\left|c\right|^2\right)  $$
Now, we use the following facts
\begin{itemize}
\item every $f\in {\rm C}_{0,2\pi}^\infty([0,2\pi])$  satisfies the condition
$f(2\pi)=e^{i\pi\epsilon/2}f(0)$
and therefore belongs in the domain of each of the operators
$(-i)^n\frac{d^n}{d x^n}$.

\item the set of functions
\begin{equation}
\sum_{k=n_0}^{n_1}a_k e^{i(k+\epsilon)x}, \ \ \ n_0,n_1\in\Z, a_k\in\C \label{finfou}
\end{equation}
is dense in the domain of each of the operators $(-i)^n\frac{d^n}{d
x^n}$ in graph norm ( the set is spanned by the  eigenfunctions).

\item in the consequence, every  $f\in {\rm C}_{0,2\pi}^\infty([0,2\pi])$
  belongs in the closure
of space (\ref{finfou}) in the right-hand-side norm (\ref{norms}).

\item the graph norm of the operator ${H''}_{\rm APS}$ (the left-hand-side
of (\ref{norms})) is weaker  then the right-hand-side norm, hence
every
 $f\in {\rm C}_{0,2\pi}^\infty([0,2\pi])$ belongs in the closure
of space (\ref{finfou}) in the right-hand-side norm (\ref{norms}).
\end{itemize}
That completes the proof.

\end{proof}

In the next subsection we will see that the set of functions
C$^\infty_{0,2\pi}([0,2\pi])$ is in fact sufficient for our
purposes because  the restriction of the operator ${H''}_{\rm
APS}$ to that set  is an essentially self adjoint operator. (It
follows, in particular, that $D({H''}_{\rm APS})$ is the closure
of C$^\infty_{0,2\pi}([0,2\pi])$ in the graph norm defined by
${H''}_{\rm APS}$.)

\section{Unitary equivalence with particle on a real line.} The
last step of our construction is a unitary map
\[
W:L^2\left(\left(-\infty,\ \infty\right)\right)\ \rightarrow\ L^2([0,2\pi]).
\]

Consider the following function
\[
y(x)=\frac{1}{2}\ln\tg\frac{x}{4}
\]
It transforms bijectively interval $[0,2\pi]$ onto $\R$.
Define $W$ to act on $f\in {\rm C}^\infty_0(\R)$ in the following way
\[
W(f)(x)\ =\ \sqrt{\frac{d y}{d x}} f(y(x))
\]
We prove
\begin{lm}
The operator ${H''}_{\rm APS}$ (\ref{Hbis}) defined in the domain
C$^\infty_{0,2\pi}([0,2\pi])$ (\ref{thedomain}) in the Hilbert
space $L^2([0,2\pi])$ is mapped by the unitary map $W^{-1}$ into
the following operator defined in the domain C$^\infty_0(\R)$ in
the Hilbert space $L^2(\R)$,
\[
W^{-1}{H''}_{\rm APS} W\ =\ -12\pi G \left(\frac{d^2}{d y^2} +
\frac{\alpha+1}{4\cosh^2(2\hat{y})}\right).
\]
\end{lm}
\begin{proof}
First, we compute the  image under $W$ of the operator $i \frac{d}{d y}$
defined in the domain C$^\infty_{0}(\R)$ in the Hilbert space
$L^2(-\infty,\infty)$, and find the following result,
\[
W i\frac{d}{d y}W^{-1}\ =\
2i\sin(\frac{\hat{x}}{2})\frac{d}{d x}+\frac{i}{2}\cos(\frac{\hat{x}}{2})
\]
defined in the domain C$^\infty_{0,2\pi}([0,2\pi])$. We have used  the formula
\[
\frac{d y}{d x}=\frac{1}{2\sin(\frac{x}{2})}.
\]
Next, we write ${H''}_{\rm APS}$ in the following form
\begin{equation}
{H''}_{\rm APS}\ =\ - 3\pi G \left( \left( 4
\sin{\frac{\hat{x}}{2}}\frac{d}{dx} +
\cos(\frac{\hat{x}}{2})\right)^2 \ +\
(\alpha+1)\sin^2(\frac{\hat{x}}{2}) \right).
\end{equation}
Finally, we note, that
\[
\sin(x(y))=\frac{2\tg(x/4)}{1+\tg^2(x/4)}=\frac{2e^{2y}}{1+e^{4y}}\ =\
\frac{1}{\cosh(2y)}.
\]
\end{proof}

\section{Remarks and outlook}  The results of this paper were presented and
discussed in Section 2. The established self-adjointness in the
$\Lambda\le 0$ case explains why the numerical computation of the
evolution of the quantum state of the Universe made in
\cite{APS,ACS} went so well. On the other hand, we were not able to
generalize our proof to the case of $\Lambda>0$. Also, some new
results \cite{Tomek}  suggest a conjecture that the operator in
question may have more then one self-adjoint extension when
$\Lambda>0$.

Our results apply to the generalization (\ref{HAPS}) of the operator
$B^{-1}(\hat{v})\hat{C}_{\rm gr}$ derived in \cite{APS}. The value
$\alpha=5/9$  corresponds to the operator
$B^{-1}(\hat{v})\hat{C}_{\rm gr}$. However, if $\alpha=0$, then the
corresponding function $\tilde{B}$ is just the inverse volume
operator (modulo a constant factor). That case is being currently
studied closer by Ashtekar, Corichi and Singh \cite{ACS}
independently of our work. Our result ensures that in that case the
operator $H_{\rm APS}$ is essentially self-adjoint as well provided
$\Lambda\le 0$.
\bigskip

In this paper we focus on the flat FRW model. However the arguments
used in the $\Lambda<0$ case can be easily generalized to the $k=-1$
and $k=1$ cases to prove the self-adjoitness and discreteness of the
spectrum of the $B(\hat{v})^{-1}\hat{C}_{\rm gr}$ operator.

The summary and new extensions of the results achieved in
\cite{k=1}, \cite{k=-1} and in the current paper,  as well as a
discussion of the relevance for Loop Quant Gravity \cite{LQG} will
be discussed in \cite{kls2}.
\bigskip

 \noindent{\bf Acknowledgements} {We thank
Abhay Ashtekar, Tomasz Paw{\l}owski, Parampreet Singh for
stimulating discussions. The work was partially supported by the
Polish Ministerstwo Nauki i Szkolnictwa Wyzszego grant  1 P03B 075
29.

\end{document}